\documentclass[prb,aps,twocolumn,superscriptaddress,showpacs]{revtex4-1}

\usepackage{graphicx}% Include figure files
\usepackage{epstopdf}
\usepackage[colorlinks=true,linkcolor=black, citecolor=blue, urlcolor=blue,
    unicode=true]{hyperref}

\begin{document}

\title{From mean-field localized magnetism to itinerant spin fluctuations in the ``Non-metallic metal" - FeCrAs}

\author{K.~W. Plumb}
\altaffiliation[Permanent address: ]{Department of Physics, Brown University, Providence RI, 02902}
\affiliation{Department of Physics, University of Toronto, Toronto, Ontario, M5S 1A7, Canada}
\author{C. Stock}
\affiliation{School of Physics and Astronomy and Centre for Science at Extreme Conditions, University of Edinburgh, Edinburgh EH9 3FD, UK}
\author{J. A. Rodriguez-Rivera}
\affiliation{NIST Center for Neutron Research, National Institute of Standards and Technology, 100 Bureau Dr., Gaithersburg, MD 20899}
\author{J.-P. Castellan}
\affiliation{Laboratoire L\'eon Brillouin, CEA-CNRS UMR 12, 91191 Gif-sur-Yvette Cedex, France}
\affiliation{Institut f{\"u}r Festk{\"o}rperphysik, Karlsruher Institut für Technologie, P.O. 3640, D-76021 Karlsruhe, Germany}
\author{J.~W. Taylor}
\affiliation{Rutherford Appleton Lab, ISIS Facility, Didcot OX11 0QX, Oxon, England}
\affiliation{European Spallation Source ERIC, Odarsl\"ovsv\"agen 113, 225 92 Lund, Sweden}
\author{B. Lau}
\affiliation{Department of Physics, University of Toronto, Toronto, Ontario, M5S 1A7, Canada}
\author{W. Wu}
\affiliation{Department of Physics, University of Toronto, Toronto, Ontario, M5S 1A7, Canada}
\author{S. R. Julian}
\affiliation{Department of Physics, University of Toronto, Toronto, Ontario, M5S 1A7, Canada}
\affiliation{Canadian Institute for Advanced Research, Quantum Materials
    Program, MaRs Centre, West Tower, Suite 505, Toronto, Ontario, M5G 1M1,
    Canada}
\author{Young-June Kim}
\affiliation{Department of Physics, University of Toronto, Toronto, Ontario, M5S 1A7, Canada}
\date{\today}

\begin{abstract}

FeCrAs displays an unusual electrical response that is neither metallic in
character nor divergent at low temperatures, as expected for an insulating
response, and therefore it has been termed a ``nonmetal-metal".  The anomalous
resistivity occurs for temperatures below $\sim$ 900 K.  We have carried out
neutron scattering experiments on powder and single crystal samples to study
the magnetic dynamics and critical fluctuations in FeCrAs.  Magnetic neutron
diffraction measurements find Cr$^{3+}$ magnetic order setting in at
$T_{N}$=115 K$\sim$10 meV with a mean-field critical exponent. Using neutron
spectroscopy we observe gapless, high velocity, magnetic fluctuations emanating
from magnetic positions with propagation wave vector $\vec{q}_{0}=({1\over3},
{1\over 3})$, which persists up to at least 80 meV$\sim$927 K, an energy scale
much larger than $T_{N}$.   Despite the mean-field magnetic order at low
temperatures, the magnetism in FeCrAs therefore displays a response which
resembles that of itinerant magnets at high energy transfers.  We suggest that
the presence of stiff high-energy spin fluctuations extending up to a
temperature scale of $\sim$900 K is the origin of the unusual temperature
dependence of the resistivity.

\end{abstract}


\maketitle

\section{Introduction}

There is a growing list of materials which behave as neither a metal nor an
insulator.~\cite{Stewart01:73}  Recent examples of interest include underdoped
and high temperature superconducting cuprates, heavy fermions, iron based
chalcogenides~\cite{Sales09:103,Rosler11:84,Rodriguez13:88}, and
oxyselenides~\cite{Zhu10:104,McCabe14:89,Stock16:28}.  The underlying cause of
this  unconventional behavior is not understood on a general level and, as
found in at least the cuprates~\cite{Keimer15:518} and iron based
superconductors~\cite{Dai15:87}, is often complicated by several competing
structural and magnetic orders. Here, we report neutron scattering measurements
studying the magnetic fluctuations in an extreme example of this unusual
electronic behavior found in FeCrAs.

FeCrAs displays very unusual properties which have led to it being termed a
``non-metallic metal" \cite{Wu09:85,Akrap14:89}.  Thermodynamic measurements
reveal a highly enhanced Fermi liquid: the linear coefficient of specific heat
is $\gamma \sim$ 30~mJ/mole\,K$^2$, while the susceptibility is Pauli like and
quite large, leading to a Wilson ratio of approximately 4. On the other hand,
not only does the electrical resistivity $\rho(T)$ show a strong departure from
Fermi liquid $T^2$ behavior --  as $T \rightarrow 0$ K it has a {\em
    sub-linear} power law $\rho(T) \simeq \rho_\circ + A T^{0.6}$ -- but it is
also ``non-metallic" in the sense that the $A$ coefficient is {\em negative}.
That is, the resistivity rises with  decreasing temperature, but without any
evidence of a gap in the density of states. In contrast to the Kondo effect,
where such behaviour is seen only at low temperature, in FeCrAs the resistivity
has a negative slope over a huge temperature range. The $ab$-plane resistivity
rises monotonically with decreasing temperature from near 900~K down to the
lowest measured temperatures of 80 mK, while the $c$-axis resistivity has a
similar rising form interrupted only by a sharp fall just below the
antiferromagnetic ordering temperature $T_N \simeq 125$~K.  The magnitude of
the resistivity is in the range of a few hundred $\mu\Omega$\,cm, which is very
large for a metal. First principles calculations predict a carrier density of
approximately $2\times 10^{28}$ m$^{-3}$, and for this density the measured
resistivity would suggest an extremely short mean-free-path, well below one
lattice spacing.

\begin{figure}[t]
    \includegraphics[width=3.25in]{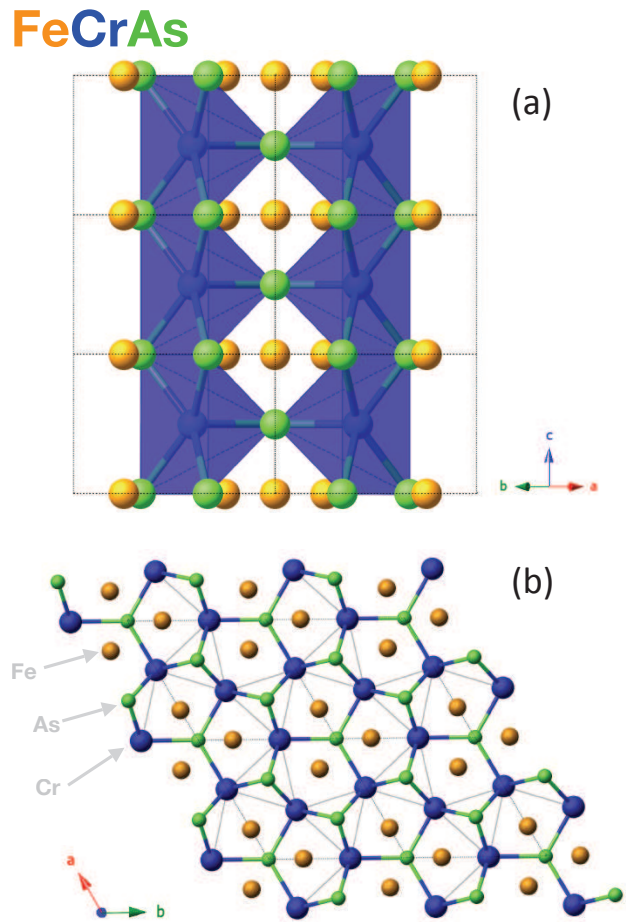}
    \caption{\label{fig:structure} (a) Crystal structure of FeCrAs illustrating
        the CrAs$_{5}$ pyramids and connectivity along the $c$-axis. (b) The
        structure projected onto the ab-plane.}
\end{figure}

FeCrAs has a hexagonal crystal structure (space group $P\overline{6}2m$ with
lattice constants $a$=6.068 \AA\  and $c$= 3.657 \AA). The Fe sublattice forms
a triangular lattice of trimers, while the Cr ions form a highly distorted
Kagome framework within the basal plane (See Figure~\ref{fig:structure}).
However, the interlayer Cr-Cr distance is relatively short (3.657 \AA),
suggesting that the interlayer hopping is substantial in this material. This is
consistent with the small resistivity anisotropy ($\rho_c/\rho_{ab} < 2$). The
Cr magnetic moments order at T$_{N} \sim$125 K forming a spin-density wave with
the ordered moments varying from 0.6 to 2.2 $\mu_{B}$.~\cite{Swainson10:88}
Given that the Cr magnetic moment measured with neutrons is proportional to
$gS$ ($g$ is electron gyromagnetic ratio and $S$ is spin quantum number), it is
likely that Cr has  valence of $3+$ (hence $S={3\over 2}$) and therefore lacks
an orbital degeneracy in pyramidal crystal field environment. In contrast to
iron based pnictides, earlier studies report that the Fe site in FeCrAs does
not carry an observable moment at any temperature. The neutron diffraction
results found no ordered moment at the Fe site.\cite{Swainson10:88} A
fluctuating Fe moment should result in some induced polarization when the Cr
sublattice orders, but this is not observed  in M\"ossbauer
spectroscopy.\cite{Rancourt:thesis} Linear spin density approximation
calculation suggests significant covalency between Fe and As, so that moment
formation is negligible (i.e. it is below the Stoner criterion).\cite{Ishida96}
All of these studies are also consistent with the suppressed Fe K$\beta^\prime$
fluorescence line observed by X-ray emission spectroscopy, which is sensitive
to any fluctuating moment down to the x-ray time scale.\cite{Gret11:84} These
combined observations provide compelling evidence that any static and dynamic
Fe moment is negligibly small in FeCrAs.

A number of theoretical studies have been devoted to understanding the strange
metallic properties of FeCrAs.  The magnetic phase diagram of the coupled Fe
trimer lattice and the distorted Kagome lattice of Cr has been mapped out
predicting magnetic order consistent with experiment.~\cite{Redpath11:xx}
Given the lack of observable static magnetic order on the Fe sublattice, a
hidden spin-liquid phase has been proposed arising from the close proximity to
a metal-insulator transition. The strong charge fluctuations associated with
this nearby critical point have been implicated as the origin of the unusual
transport properties.~\cite{Rau11:84}  An alternate explanation has been
proposed in the context of ``Hund's metals'' where large localized moments are
coupled to more itinerant electrons.~\cite{Nevi09:103,Yin11:10} There have been
only limited number of spectroscopic studies to put these theories to test.
Charge excitations have been investigated using optical spectroscopy which
revealed that the anomalous temperature dependence of resistivity was dominated
by the temperature dependence of scattering rate, rather than
carrier-concentration.\cite{Akrap14:89} In addition, they found that two Drude
components with drastically different energy scales contribute to the low
energy charge dynamics. On the other hand, the spin dynamics in FeCrAs have not
been investigated to date.

In this study, we apply neutron scattering to investigate the magnetic
properties of FeCrAs with emphasis on the static order and fluctuations
originating from the Cr$^{3+}$ sites. We first present diffraction work showing
the magnetic order associated with the propagation wave vector of
$\vec{q}_0=({1\over3}, {1\over3})$, is described with a mean-field critical
exponent. We then measured the powder averaged fluctuations showing stiff
magnetic fluctuations extending up to at least $\sim$ 80 meV, while the low
energy excitations seem to be well described with gapless spin waves emanating
from the ordering wave-vector. These results illustrate spin excitations in
FeCrAs resemble those in itinerant magnets. We further discuss the magnetic
excitation spectrum in the context of the unusual transport properties. Our
finding of a high energy scale for magnetic fluctuations suggests that magnetic
fluctuations could be responsible for the anomalous scattering that is observed
up to high temperatures, despite the N\'eel temperature occurring at much lower
temperature. Although our observations do not directly speak to the mechanism
of non-metallic and non-Fermi-liquid resistivity in the $T \rightarrow 0$ K
limit, it seems natural to hypothesize that anomalous magnetic correlations
begin to form at very high temperature in FeCrAs, and continue to evolve down
to very low temperature, somehow producing the non-metallic metal state.

\section{Experiment Details}

The powder samples of FeCrAs were prepared by melting high purity Fe, Cr, and
As in stoichiometric ratios following Ref. \onlinecite{Katsuraki66:21}. A small
single crystal (with mass 25 mg) was also produced by slow cooling from the stoichiometric
melt. The single crystal used here was from the same batch as those used in
earlier transport and thermodynamic studies discussed in Ref. \onlinecite{Wu09:85}.

High energy inelastic neutron scattering measurements on powder samples were performed
using the MARI direct geometry chopper spectrometer (ISIS, Didcot).
Measurements were performed with incident energies of E$_{i}$=150~meV, and
300~meV that were selected using the ``relaxed'' Fermi chopper spinning at
f=300~Hz, and 450~Hz respectively with the data being collected in a time of
flight mode.   Details of the background subtraction are provided below.
Single crystal spectroscopy measurements were not successful owing to the small
sample size.

Further higher resolution neutron spectroscopy measurements were performed on
the MACS cold triple-axis spectrometer (NIST, Gaithersburg).   Instrument and
design concepts can be found elsewhere.~\cite{Rodriguez08:19,Broholm96:369}
Data was collected by measuring momentum space cuts at constant energy
transfers by fixing the final energy at E$_{f}$=2.4 meV
using the 20 double-bounce PG(002) analyzing crystals and detectors and varying
the incident energy defined by a double-focused PG(002) monochromator.    Each
detector channel was collimated using 90$'$ Soller slits before the analyzing
crystal and a cooled Be filter was placed before the analyzing crystals.
Maps of the spin excitations as a function of energy transfer were then
constructed from a series of constant energy scans at different energy
transfers.  All of the data has been corrected for the $\lambda/2$
contamination of the incident-beam monitor and an empty cryostat measurement
was used to estimate the background.

Single crystal magnetic neutron diffraction measurements were performed on the
1T1 thermal triple axis spectrometer (LLB, Saclay) utilizing an open
collimation sequence, double focusing monochromater and vertically focusing
analyzer. The crystal was aligned in the (HK0) scattering plane of the
hexagonal unit cell for the duration of the experiment.

\section{Results}

\subsection{Magnetic order from neutron diffraction}
\begin{figure}[t]
    \includegraphics[]{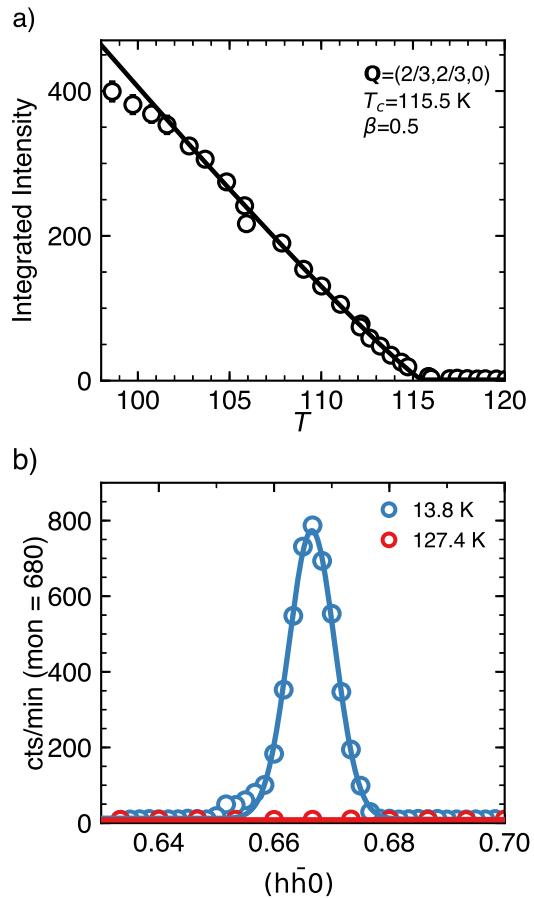}
    \caption{\label{fig:order_param} (a) Integrated intensity of the
        (2/3,2/3,0) magnetic Bragg peak measured on 1T-1 (LLB), solid line is a
        fit to $\left(1-T/T_N\right)^{2\beta}$ with $T_N\!=\!115.5\pm 0.5$~K
        and $\beta\!=\! 0.54\!\pm\!0.05$.  (b) Transverse scans through the
        magnetic Bragg peak at (2/3,2/3,0).}
\end{figure}

Neutron diffraction characterizing the magnetic order is presented in
Fig.~\ref{fig:order_param}. The resolution limited magnetic Bragg peaks in
Fig.~\ref{fig:order_param} (b) confirm the presence of long-range magnetic
order with a $({1\over 3}, {1\over 3})$ propagation vector as observed in
previous powder diffraction measurements.\cite{Swainson10:88}  The integrated
neutron scattering intensity which is proportional to the squared magnetic
order parameter is plotted in Fig.~\ref{fig:order_param} (a). We observe the
onset of magnetic Bragg intensity at $T_N\!=\!115.5(5)$~K, a temperature
significantly lower than the $T_N\!=\!125$~K N\'eel temperature extracted from
resistivity and magnetic susceptibility measurements on the same sample
(Ref.~\onlinecite{Wu09:85}). The value of $T_N$ in FeCrAs is known to vary across
different samples between 100 and 125~K depending on the synthesis conditions
and sample quality. Those samples with a higher T$_N$ are observed to have a
splitting of field cooled and zero field cooled magnetic susceptibility at
lower temperatures and the highest quality samples are associated with the
highest $T_N$.\cite{Wu11:170} However, neutron diffraction and magnetic
susceptibility measurements were performed on the same sample so the origin of
this discrepancy is presently not clear.

In a mean field approximation, for localized magnetism, the critical
temperature is related to the magnetic exchange interaction via the relation,
\begin{eqnarray}
k_B T_{N} \sim k_{B}\Theta_{CW}={2 \over 3} S (S+1)zJ,
\label{Curie_Weiss} \nonumber
\end{eqnarray}
where $S$ is the spin value, presumably ${3 \over 2}$ for Cr$^{3+}$, and $J$ is
the average exchange constant with $z$ representing the number of nearest
neighbors. The FeCrAs magnetic structure is highly
frustrated~\cite{Florez15:27} potentially making the sum over neighbors quite
complicated.  However, this expression does allow us to obtain an estimate of
the mean-field spin-wave velocity of $zSJa \sim 3 k_B T_N a/2(S+1)\sim
20$~meV$\cdot$\AA\, if we assume local spin moments.  We compare this energy
scale to the measured spin fluctuations below.

A fit of the temperature dependent integrated neutron intensity to a power law
near $T_{c}$ finds the mean-field critical exponent $\beta$=0.54 $\pm$ 0.05.
This differs from the critical exponent of $\beta \sim $0.25 found in iron based
langasite~\cite{Stock11:83} and other two-dimensional triangular
magnets~\cite{Kawamura88:63}. The fluctuations critical to magnetic order in
FeCrAs also differ from iron based pnictides and chalcogenides which broadly
display Ising universality class
behavior.~\cite{Wilson10:81,Wilson09:79,Pajerowski13:87,Stock16:28} However,
the mean-field critical exponent is expected for an itinerant ferromagnetic
transition. For example, Moriya's spin-fluctuation theory
predicts the temperature dependence of $M \sim
(1-T/T_c)^{1/2}$.\cite{Mohn_book}

\subsection{Magnetic dynamics from inelastic neutron scattering}
\begin{figure}[t]
\includegraphics[width=8.5cm] {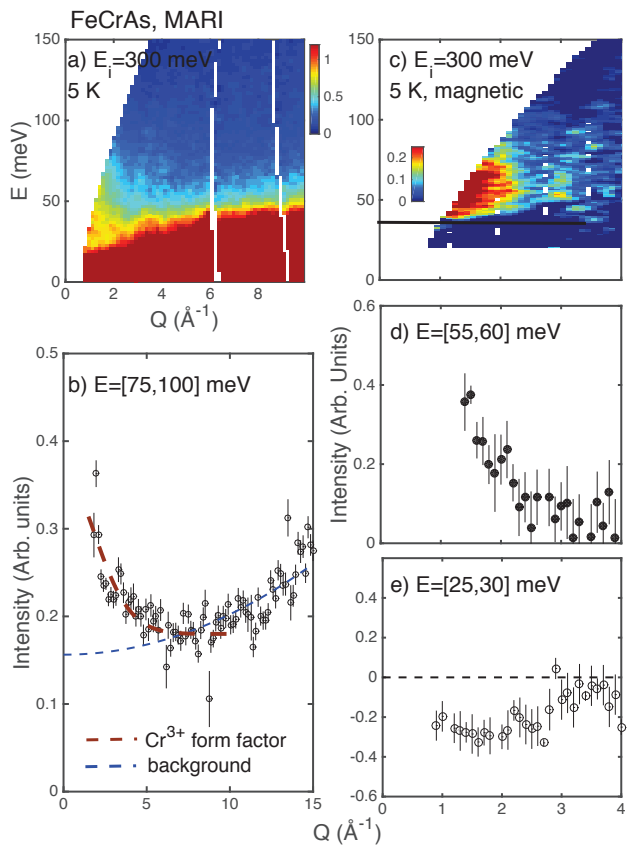}
 \caption{\label{exp} (a) Powder-averaged inelastic neutron spectrum in FeCrAs
     taken on MARI. The intensity between 75~meV and 100~meV is integrated and
     plotted as a momentum cut in panel (b). The blue dashed line is
     an estimate of the background from extrapolating from large momentum
    transfers as described in the text and the dark red curve is the scale
    magnetic Cr$^{3+}$ form factor.  (c) Illustrates the same data as in panel
    (a), but with the background removed.  The solid line at $E$=40~meV shows
    where the background subtraction fails due to strong and highly structured
    in momentum phonon scattering. Constant energy cuts from this panel are
    plotted in (d) for the energy interval $E$=[55,60] meV, and (e) for the
    energy interval $E$=[25,30] meV. }
\end{figure}

We now discuss the magnetic dynamics as measured by inelastic neutron
scattering. Figure \ref{exp} illustrates the high-energy spectroscopy
measurements performed on the MARI chopper spectrometer.  Panel (a) displays a
powder-averaged energy-momentum map at 5~K showing the presence of scattering
at low momentum transfers above $\sim$ 50~meV which decays rapidly with $Q$.
The white region corresponds to where no data could be taken due to kinematic
constraints of neutron scattering imposed by a minimum scattering angle of
$2\theta \sim 3 ^{\circ}$.  Panel (b) shows a constant energy cut illustrating
the presence of two components to the scattering: one rapidly decaying with
momentum, indicative of magnetic fluctuations and well described by the
Cr$^{3+}$ form factor, and the other slowly increasing at large momentum
transfers, characteristic of a phonon contribution. To extract magnetic
fluctuations at high energy transfers, we relied on the fact that the magnetic
scattering is confined to small momentum transfers and decays with increasing
$Q$ while the phonon background increases with $Q^2$.

We have separated the two components by fitting the high angle detector
intensity (where magnetic scattering is expected to be negligibly weak) to
$I_{BG}=B_{0}+B_{1}Q^{2}$  and extrapolating to small momentum transfers.  An
example of this analysis is illustrated by the dashed curves in Fig.~\ref{exp}
(b) which show a cut integrated over energies between 75 and 100 meV. The
dashed lines in (b) show an estimate of the background based on a fit to the
high angle detectors and also the Cr$^{3+}$ form factor scaled by a constant
factor to agree with the low-$Q$ momentum dependence. The result of applying
this analysis to each energy transfer and subtracting the high-$Q$ background
is shown by the false color image in panel (c).  Individual cuts integrating
over $E$=[55,60] meV and $E$=[25,30] meV are plotted in panels (d) and (e).
The analysis successfully extracts magnetic intensity for energy transfers
above $\sim$ 45 meV, but failed to separate out the magnetic and phonon
contribution at lower energy transfers resulting in an over subtraction of
intensity.  This is seen in the false color image in panel (c) and further
displayed through constant energy cuts in panels (d) and (e). While the
background subtraction works at large energy transfers as shown in panel (d),
the assumptions behind this background correction break down for low-energy
transfers, where the phonon scattering becomes intense and highly structured in
momentum as shown in panel (e).  Therefore, we have removed the region below
20~meV from the plots.  We note that this technique for background subtraction
has been successfully applied previously to studying high energy $d-d$
transitions in NiO and CoO.~\cite{Kim11:84, Cowley13:88}  It was also applied
to extract the magnetic fluctuations in $\alpha$-NaMnO$_{2}$.\cite{NaMnO2}  In
all of these cases the analysis was only applied to a region in momentum-energy
where the powder averaged phonon contribution was small and unstructured in
$Q$.

\begin{figure}[t]
\includegraphics[width=9.5cm] {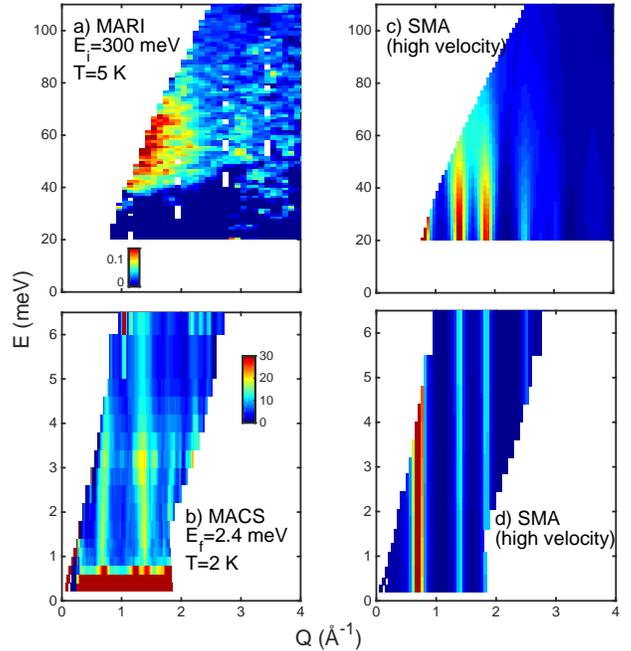}
\caption{\label{constQ} The powder averaged magnetic response at 5 K in FeCrAs
 measured with (a) $E_{i}$=300 meV (MARI, ISIS) and (b) $E_{f}$=2.4 meV
 (MACS, NIST).  The variation in pixel size as a function of energy transfer in
 the MACS data, panel (b), is due to the difference in the way the data is
 collected.  MARI data was collected in a time of flight configuration while
 MACS is a triple-axis which each energy transfer corresponding to a different
 constant energy scan.  (c-d) The powder averaged heuristic parametrization
 based on the single mode approximation (SMA) discussed in the text.  The
 calculation was done assuming two dimensional linear spin-waves with a
 velocity of 200 meV $\cdot$ \AA. }
\end{figure}

\begin{figure}[t]
\includegraphics[width=8.5cm] {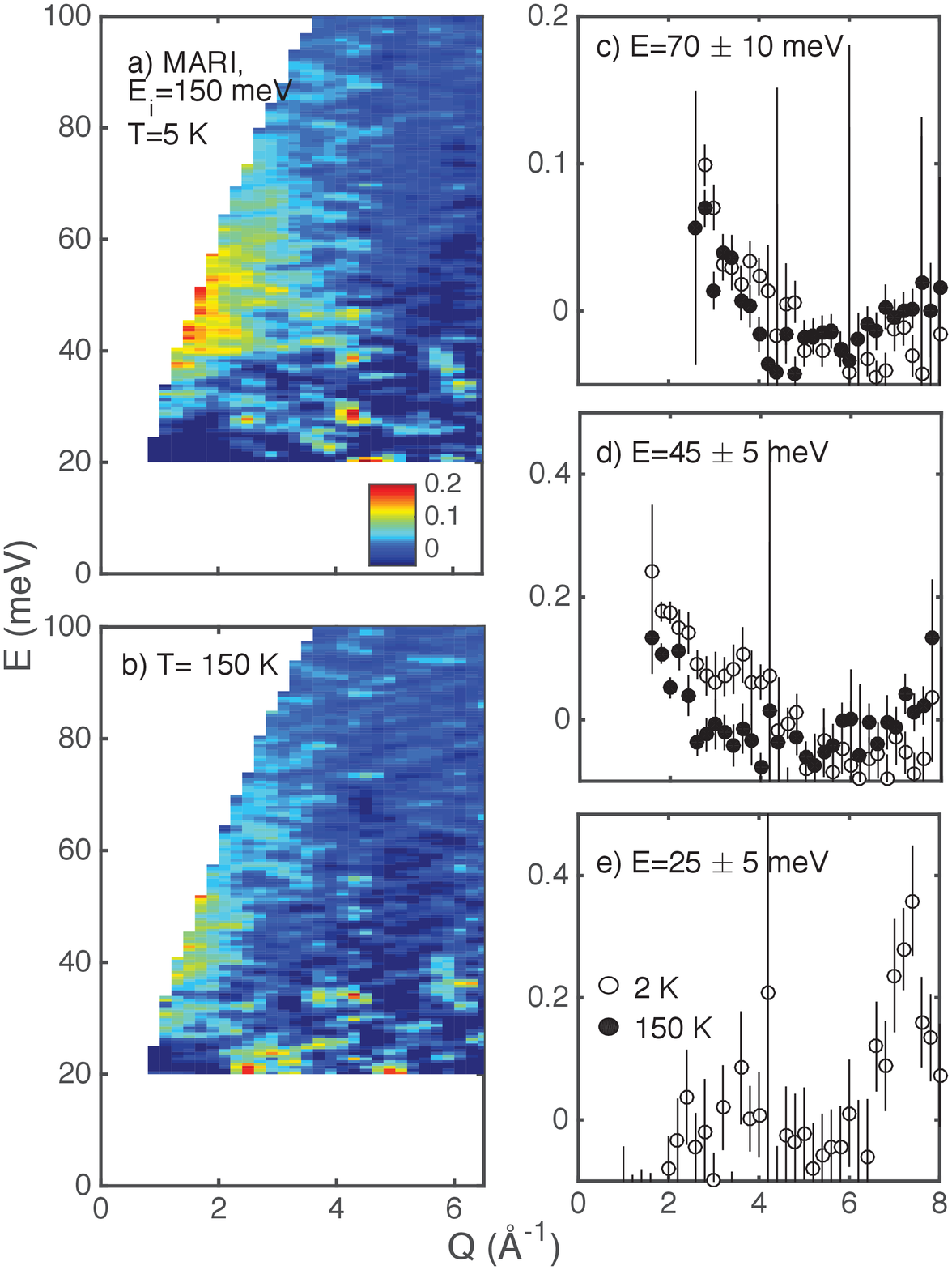}
\caption{\label{temp} The powder averaged magnetic response measured with
    $E_{i}$=150 meV at (a) T=5 K and (b) 150 K.  (c-d) Constant energy cuts
    illustrating the decay of magnetic intensity with momentum transfer.  Panel
    (e) illustrates an energy cut at 25 meV where phonon scattering prevents a
    reliable subtraction of the background.}
\end{figure}

Given the failure to extract reliable magnetic scattering below $\sim$ 40 meV
using the MARI direct geometry spectrometer we have used the MACS cold
triple-axis spectrometer with a low fixed $E_{f}$=2.4 meV to investigate the
magnetic response at low energy transfers. This configuration kinematically
affords access to low momentum transfers where phonon scattering is expected to
be negligible. The background corrected data from MACS is compared against the
high-energy magnetic response extracted used in MARI in Fig.~\ref{constQ}
(a-b). Steeply dispersing magnetic fluctuations are observable at low energies
below $\sim$ 6~meV, emanating from $Q$ positions which correspond to the
propagation vector of $\vec{q}_{0} = ({1\over3}, {1\over3})$. Further magnetic
fluctuations are observable above $\sim$ 40 meV using MARI. Data between these
two energy ranges, bridging the MACS and MARI data sets, could not be reliably
extracted, as discussed above, due to both kinematic constraints of neutron
scattering and also the substantial phonon background over this energy range
highlighted in Fig. \ref{exp} panel (c).

The MACS data in Fig.~\ref{constQ} (b) reveal additional weak magnetic scattering
near 3~meV  suggestive of a second low-frequency magnetic mode.  It is possible
that this mode is the second transverse mode (magnon) with a gap of about 3~meV
resulting from a weak easy plane anisotropy. Another possibility is that this
intensity arises from a longitudinal mode, similar to what has been found in
other metallic magnets.\cite{Endoh06} Experiments using single crystal
samples are necessary to address the nature of these low energy modes.  The high velocity, or stiff, spin
excitations extend up to 6~meV beyond which they are outside of the observation
window on MACS.  The fact that these excitations form steep rods in $Q$ seen in
the MACS data allows us to speculate that they link to the high-energy response
observed on MARI.  We discuss this point below by applying a parameterization,
illustrated in Fig.  \ref{constQ} (c-d), based on the first moment sum rule.

Above, we have relied on the momentum dependence to extract the magnetic
intensity.  To further confirm the magnetic origin of the low-angle response,
we have measured the magnetic fluctuations at higher temperature shown in
Fig.~\ref{temp} which plots the extracted magnetic scattering with
$E_{i}$=150~meV at 5~K and 150~K, below and above $T_{N}$ respectively.
Background corrected false color maps at these two temperatures are shown in
panels (a-b) with constant energy cuts shown in panels (c-d).  While the
low-$Q$ excitations are still present at high temperatures, indicative of a
large underlying energy scale, a decrease in the scattering confirms the
magnetic origin of this scattering present at small momentum transfers.

\subsection{Parameterization in terms of high velocity damped spin waves}

The two data sets from time of flight and triple-axis spectroscopy show
magnetic excitations at high and low energy regime quite clearly; however, we
note that it is difficult to measure magnetic excitations in the intermediate
energy regime connecting these two data sets because of strong phonon
scattering. To illustrate a consistent link between the low and high energy
data sets, we have parameterized the spin fluctuations by high velocity damped
spin-waves from the magnetic $=({1 \over 3},  {1\over 3})$ positions.  We have
simulated the scattering using the following form motivated by the
Hohenberg-Brinkmann first moment sum rule applied in the case of a dominant
single mode, known as the single mode approximation.~\cite{Hohenberg74:10}
This approach has been applied to low-dimensional organic magnets (Refs.
\onlinecite{Hong06:74,Stone01:64}) and the form reflects that used to describe
magnetic excitations in powder samples of triangular magnets (Refs.
\onlinecite{NaMnO2,Wheeler09:79}).

\begin{eqnarray}
S(\vec{Q},E)\propto {1\over \epsilon(\vec{Q})} \gamma(\vec{Q})  f^{2} (Q) \delta (E-\epsilon(\vec{Q})),
\label{equation_rho} \nonumber
\end{eqnarray}

\noindent where $\gamma(\vec{Q})$ is a geometric term chosen to peak at the
Bragg positions with propagation vector $({1\over 3} , {1\over 3})$, $f^{2}(Q)$
is the magnetic form factor for Cr$^{3+}$, $\delta (E-\epsilon(\vec{Q}))$ is an
energy conserving delta function, and $\epsilon(\vec{Q})$ is the dispersion
relation for the spin excitations. We only consider Cr$^{3+}$ moments here
because the iron moment is negligibly small as discussed above. Given that the
scattering is concentrated at low momentum transfers and a large portion is
kinematically inaccessible, we are not able to derive an accurate measure of
the total integrated intensity for comparison to sum rules of neutron
scattering.  For the calculations shown here, we have taken the spin wave
dispersion to be two dimensional (within the $a-b$ plane) and also be linear
given that no upper band is observed.  Powder averaging was done using a finite
grid of 10$^{4}$ points and summed at each momentum and energy transfer.
We note due to powder averaging it is difficult to make any conclusions from
the data regarding any continuum scattering that may exist owing to
longitudinal spin fluctuations as observed in other itinerant
systems~\cite{Stock15:114}.  As displayed in Fig.  \ref{constQ} panel (d),  the
combination of powder averaging results in scattering over an extended range in
momentum transfer.  Given kinematics associated with the $({1\over 3} , {1\over
    3})$ type order, we are not able to draw any conclusions about possible
ferromagnetic fluctuations that may exist near $Q$=0.

The results of this calculation using a linear and three dimensional spin-wave velocity of $\hbar c$=
200 meV $\cdot$ \AA\ and performing the powder average are shown in Fig.
\ref{constQ}(c-d).  The calculation confirms that the two experimental data
sets presented in Fig.~\ref{constQ} (a-b) can be consistently understood in
terms of high velocity spin-waves emanating from the (${1\over 3}$, ${1\over
    3}$) positions. As seen in Fig. \ref{constQ}(c), the magnetic form factor
ensures that the magnetic scattering is suppressed at large momentum transfers.
The value used in this calculation, $\hbar c$= 200 meV $\cdot$ \AA\, should be
considered as a lower bound of the spin wave velocity. The steep velocity
ensures magnetic scattering is confined to low scattering angles as observed
experimentally which are eventually completely masked at high energy transfers
by kinematic constraints of neutron scattering.  One thing that is not clear in
this analysis is the highest energy scale of the steeply dispersing magnetic
excitations.  Our measurements do not reveal a high energy peak in the powder
averaged spectra that would result from an enhanced density of states for zone
boundary spin waves and instead we observe an apparent high energy continuum.
This may be attributed to either a combination of kinematic constraints and the
magnetic form factor, or possibly to strong damping of the highest energy
magnetic excitations that results from coupling to conduction electrons. The
latter case occurs in classic itinerant magnets.~\cite{Ishikawa77:16}  We note that our model of three dimensional spin waves emanating from magnetic $({1\over 3} , {1\over 3})$ does not capture the momentum dependent intensity of the spin excitations at low energy transfers measured on MACS (Fig. \ref{constQ} (b)).  We speculate that such modulation with momentum originates from a more complex momentum dependence not captured in our analysis originating from unusual magnetic structure.  To refine a model to capture this, single crystal data is required.

Our parameterization of the data in terms of three dimensional and high velocity spin waves emanating from $\vec{q}_{0}$=(1/3, 1/3) positions is arguably the simplest model that is consistent with the three dimensional nature of the resistivity and also the structure discussed above.  However, it should be emphasized that powder averaging does mask features that would become clear in single crystals.  It is possible that the three dimensional character of the spin excitations is only present at low energies crossing over to two dimensional excitations at higher energies.  Indeed, our heuristic model does not capture the additional scattering at $\sim$ 3 meV measured on MACS which could be suggestive of such a scenario.  As an example, we point to powder averaged spin excitations in BaFe$_{2}$As$_{2}$~\cite{Ewings08:78} which did give clear ridges of scattering up to high energies while later single crystal work confirmed the two dimensional character.  Our data and parameterization does show that high velocity spin excitations are present up to unusually high energies in FeCrAs with the exact nature of the dimensionality made ambiguous from the powder averaging.

\section{Discussion}
Our neutron diffraction measurements (with resolution $\sim$ 2~meV) show that
the magnetic order sets in around $T_N=115$~K in FeCrAs, and the sublattice
magnetization is well described with mean-field critical exponent of
$\beta=1/2$. The inelastic neutron scattering measurements show that the
low-energy spin excitations of FeCrAs are well-defined gapless spin-waves
extending up to $\sim$ 6 meV.  The spin excitations are observed up to high
energy transfers of at least 80~meV.  Such a large energy scale of these spin
excitations indicate an underlying magnetic energy scale that is significantly
larger than that estimated from local moment molecular field model (See
Sec.~IIIA).

These observations are quite reminiscent of the spin excitations observed in
chromium metal. The incommensurate spin-density wave order in Cr is considered
as a textbook example of magnetic order driven by the Fermi surface
nesting.\cite{Fawcett88} Its spin excitation spectrum has been the subject of
intense investigation both theoretically and
experimentally.\cite{Fedders66,Walker76,Fishman96,Fincher79,Lorenzo94,Endoh06}
Experimentally, spin-wave like excitations with very steep dispersion have been
observed by inelastic neutron scattering
measurements.\cite{Fincher79,Lorenzo94,Endoh06} Theoretical studies showed that
the transverse spin fluctuations in the long wavelength limit can be described
by spin-wave modes even for this type of itinerant
systems.\cite{Fedders66,Walker76,Fishman96} That is, $\omega=c|q|$, but the
spin-wave velocity is given by $c=v_F/\sqrt{3}$, where $v_F$ is the Fermi
velocity which originates from charge physics and therefore is much larger than
typical spin-wave velocity observed in a localized spin model.  In addition to
the transverse spin waves, a longitudinal mode is allowed and, in fact, has
been observed to be quite strong.\cite{Lorenzo94} The observed spin wave
velocity is a weighted combination of transverse and longitudinal modes and so
can differ significantly from  $c=v_F/\sqrt{3}$. In Cr, the longitudinal
fluctuation renormalizes the apparent spin-wave velocity down \cite{Endoh06}
and  the apparent spin-wave velocity $\hbar c$(Cr)  is given by $\hbar c$(Cr)$
\sim \hbar \sqrt{c_L c_T} \sim 1000$~meV $\cdot$ \AA\, where $c_L$ and $c_T$
denote longitudinal and transverse velocity of Cr.

Since the carrier density in FeCrAs is known from the first principles
calculation ($n=2 \times 10^{28}$~m$^{-3}$), we can estimate $v_F/\sqrt{3} \sim
4000$~meV \AA. Although this value is much larger than the spin wave velocity
used in Fig.~\ref{constQ}, we do not consider this as significant numerical
discrepancy. First, the spin-wave velocity used in Fig.~\ref{constQ} is just a
lower bound, and the data will be still adequately described with a larger
value of $c$. Second, the Sommerfeld coefficient of 30~mJ/mol K$^2$ suggests a
large renormalization of the bare Fermi velocity. Finally, longitudinal
magnetic excitations are expected to reduce the apparent spin-wave velocity. Of
course a calculation based on the real band structure would be necessary to
obtain a more quantitative comparison between itinerant theory and experiment.

We would like to point out that there is a growing list of materials which
display low-energy localized excitations but itinerant fluctuations at higher
energy transfers.
Fe$_{1+x}$Te~\cite{Fruchart75:10,Koz13:88,Okada09:78,Rodriguez11:84,Yu2018} has
been found to have localized transverse fluctuations at low-energies which
cross over to high energy fluctuations resembling more itinerant
fluctuations.~\cite{Stock14:90,Stock17:95}  CeRhIn$_{5}$ shows well defined
localized spin waves which breakdown into a multiparticle
continuum.~\cite{Stock15:114}  YBa$_{2}$Cu$_{3}$O$_{6+x}$ similarly displays
localized low-energy fluctuations but itinerant fluctuations at high
energies.~\cite{Stock07:75,Stock10:82} However, unlike above materials, FeCrAs
is far from the quasi-two-dimensional limit. The observed weak resistivity
anisotropy is a strong indicator of this, with an additional support provide by
our observation of mean-field critical exponent.  In the parent compounds of
iron or copper based superconductors, the critical behavior is usually governed
by strong 2D fluctuations, giving rise to critical exponents in the range of $\beta \sim 0.2 - 0.3$, much smaller than the observed mean-field exponent.~\cite{Bramwell1993,Wilson10:81,Wilson09:79,Pajerowski13:87,Stock16:28}

We now discuss the relation between the spin excitations and the unusual
response measured in resistivity.  The data shows fast spin excitations at low
momentum transfers.  The MACS data illustrate that the excitations are
originating from finite-$Q$, but extend up to high energy transfers.  A central
question in FeCrAs is the origin of the unusual metallic properties with the
resistivity increasing in a power-law fashion from 600 K. The resistivity from
spin fluctuations has been suggested to have the following form in the context
of work done on the cuprate superconductors.~\cite{Keimer91:67,Moriya90:59}

\begin{eqnarray}
\rho(T) \propto T \int_{-\infty}^{\infty} {E \over T} d\left({E \over T} \right) {{e^{E/T}} \over {(e^{E/T}-1)^{2}}}\int d^{3}q \chi '' (\vec{q},E).
\label{equation_rho} \nonumber
\end{eqnarray}

\noindent Given that the neutron scattering cross section $I(Q, E)\propto S(Q,
E)={1\over \pi} [n(E)+1]\chi ''(Q, E)$, an energy independent $\int d^{3}q \chi
'' (\vec{q},E)$ would result in a resistivity which has linear temperature
dependence.  If, however, this local susceptibility integral term decreased
slowly with increasing temperature, then a temperature independent resistivity
may be explained.  While the kinematic constraints of our experiment preclude
measurement of the temperature dependent local susceptibility, we do observe
only a  weak decrease of high energy magnetic intensity with increasing
temperature implying that the associated change in local susceptibility is
small.  The large energy scale of the fluctuations inevitably will affect the
resistivity over a very broad temperature scale.

The measurements above find two results in the context of the dynamics; first
that the magnetic excitations are gapless down to the energy scale set by the
$\sim$ 0.5 meV resolution of MACS; and second, the high energy scale
fluctuations are present at high temperatures above T$_{N}$.  The
large energy scale and gapless nature of the spin fluctuations may provide an
explanation for the unusual transport response.  A similar coupling between
spin fluctuations and the electron response was suggested in Fe$_{1+x}$Te which
also display little change in the resistivity over a broad range in
temperature.~\cite{Liu09:80}  Indeed, only when the magnetic fluctuations
become gapped in Fe$_{1+x}$Te does the resistivity drop and the two can be
correlated using the relation above.~\cite{Rodriguez13:88}  FeCrAs may
represent an extreme example with gapless spin excitations that extend up to at
least 80 meV ($\sim$ 926 K).

In summary, we studied critical behavior of the magnetic order parameter near
the Neel transition in FeCrAs, and observed that the temperature dependence of
the magnetic order parameter is described with the mean-field critical
exponent. Our neutron spectroscopy measurements reveal high velocity gapless
spin wave excitations which extend up to at least $\sim$ 80 meV, which
resembles spin excitations in itinerant magnets. We suggest that coupling
between this broad-band spin fluctuations is the origin of the unusual
resistivity measured in this ``nonmetal-metal".

\begin{acknowledgements}
We acknowledge the support of the Natural Sciences and Engineering Research Council of Canada (NSERC), Canada Foundation for Innovation (CFI), and Ontario Innovation Trust (OIT).
This work was supported by the Carnegie Trust for the Universities of Scotland, the Royal Society, and the Engineering and Physical Sciences Research Council (EPSRC).
\end{acknowledgements}

\end{document}